\documentclass{aa}

\usepackage[]{graphicx}

\begin{document}
\input{psfig.sty}

\title{Faint dwarf spheroidals in the Fornax Cluster:}
\subtitle{A flat luminosity function
}

\titlerunning{Faint dwarf spheroidals in the Fornax Cluster}

\author {Michael Hilker \inst{1} \and Steffen Mieske \inst{1,2} \and Leopoldo
Infante \inst{2}
}

\authorrunning{M.~Hilker et al.}

\offprints {mhilker@astro.uni-bonn.de}

\institute{
Sternwarte der Universit\"at Bonn, Auf dem H\"ugel 71, 53121 Bonn, Germany
\and
Departamento de Astronom\'\i a y Astrof\'\i sica, P.~Universidad Cat\'olica,
Casilla 104, Santiago 22, Chile
}

\date {Received 04 November 2002/ Accepted 01 December 2002}

\abstract{We have discovered $\simeq70$ very faint dwarf galaxies in the 
Fornax Cluster. These 
dSphs candidates follow the same magnitude-surface brightness 
relation as their counterparts in the Local Group, and even extend it to
fainter limits. The faintest dSph candidate in our sample has
an absolute magnitude of $M_V\simeq-8.8$ mag and a central surface
brightness of $\mu_V\simeq27$ mag/arcsec$^2$. There exists a tight
color-magnitude relation for the early-type galaxies in Fornax that 
extends from the giant to the dwarf regime.
The faint-end slope of the luminosity function of the early-type dwarfs is 
flat ($\alpha\simeq-1.1\pm0.1$), contrary to the results obtained 
by Kambas et al. (\cite{kamb00}).
}

\maketitle


\section{Introduction}

Little is known about the very faint end of the galaxy luminosity function (LF)
-- the realm of the dwarf spheroidal (dSph) galaxies. The current CDM models 
of galaxy and structure formation (e.g. Moore et al. \cite{moor98a}) suggest 
that dwarf galaxies were very abundant in the early universe and
represented the building-blocks of larger galaxies. Especially in galaxy 
clusters, large numbers of dwarf galaxies should have been formed. Different 
models result in a steep slope at the faint end of the LF: $-1.4 \leq d 
{\rm log} N/ d {\rm log} L \leq -2.0$ (e.g. Kauffmann et al. \cite{kauf94}).

Most of our knowlegde about the dwarf spheroidals comes from 
studies of Local Group examples (e.g. Mateo \cite{mate98}). Some authors
do not distinguish them from the somewhat brighter dwarf ellipticals (dEs), 
but here
we follow the definition given by Grebel (\cite{greb01}) who characterizes
them as diffuse, low surface brightness dwarfs with $M_V<-14$ mag and $\mu_V >
22$ mag arcsec$^{-2}$.
Although primarily old stellar populations, dSphs are
not the simple systems they were once thought to be -- many show evidence of
star formation in multiple bursts and traces of molecular gas (e.g. Grebel 
\cite{greb97}). Such gross
differences are very likely due to the effects of the environment these
galaxies exist in, and thus are efficient tracers of
evolutionary processes in galaxies. A large-scale study of these systems in 
nearby clusters would be essential for the understanding of the ``global'' 
picture of galaxy evolution. 

The faint end of the luminosity function of the Local Group is quite flat 
compared with CDM model predictions. A Schechter function fit gives a faint-end 
slope of $\alpha = -1.1\pm0.1$ (Pritchet \& van den Bergh \cite{prit99}).
Several CCD studies of galaxy clusters show rising numbers of low-luminosity 
dEs down to about $M_V\simeq-11$ mag with very different faint-end slopes in 
the range $-1.1>\alpha>-2.2$ (e.g. De Propris et al. \cite{depr95}; 
L\'opez-Cruz et al. \cite{lope97}). However, it is not known whether the 
galaxy luminosity function in clusters continues to $M_V=-9$,
as it does in the Local Group. In fact, the latest 
discoveries of previously unknown dSphs (e.g. Armandroff et al. 
\cite{arma99}) raises the question of how complete is the Local Group sample. 
Is there a true cutoff or one imposed solely by small number statistics? The
detection of a lower-L cutoff would add strong constraints to galactic
formation models.

The most complete investigation of the Fornax dwarf galaxies was done by
Ferguson (\cite{ferg89a}) as well as by Davies 
et al. (\cite{davie88}). As shown by Hilker et al. 
(\cite{hilk99a}) and Drinkwater et al. (\cite{drin01b}) the morphological
classification of Fornax members by Ferguson (\cite{ferg89a}) is very 
reliable and very few dEs were missed within the survey limits.

The luminosity function of the Fornax dwarf galaxies was studied by 
Ferguson \& Sandage (\cite{ferg88}). The faint end slope of the
dE/dS0 LF, fitted by a Schechter (\cite{sche76}) function up to $M_V\simeq-13$,
is quite flat ($\alpha = -1.08\pm0.10$).
However, in a recent study, Kambas et al. (\cite{kamb00}) report the 
discovery of a very large number of low surface brightness dwarfs in Fornax,
resulting in a steep faint-end slope of the LF ($\alpha \simeq -2.0$) down to 
$M_B \simeq -12$ mag.

\section{Observations and data analysis}

The observations were performed in an observing run in December 1999 with the 
100-inch du~Pont telescope at Las Campanas, Chile. The Wide-Field CCD images 
a 25$\arcmin$ diameter field, with a scale of about 0.774$\arcsec$/pixel.
14 fields in the central region of the Fornax cluster and one additional 
background field have been observed through the Johnson $VI$ filters. All 
nights were photometric throughout, and the seeing was in the range 
$1\farcs5$--$2\farcs0$. 

The CCD frames were processed with standard {\sc IRAF} routines, instrumental 
aperture magnitudes were derived using SExtractor2.1 (Bertin \& Arnouts
\cite{berti96}). Surface brightness profiles of all galaxies were measured 
with the ellipse fitting routines under the {\sc stsdas} package of {\sc IRAF}. 

In order to optimize the detection of faint resolved sources by the 
software routines, all CCD fields were first inspected carefully by eye, 
independently by the authors MH and SM, to search for low surface brightness 
objects with sizes of Local Group dSphs (core radii of 150-400 pc, 
corresponding to core diameter of $3\arcsec$ to $9\arcsec$ in 
Fornax distance). This search resulted in the detection of about 70 
previously uncatalogued dSph candidates.

The detection-sensitive parameters of SExtractor2.1 were then optimized such 
that most objects of the by-eye-catalog were detected by the program. 
Further visual inspection of the SExtractor detections within the same 
parameter space as the visual detections, added about 10\% more dSph 
candidates to the by-eye-catalog. About 10\% 
of the obvious by-eye detections could still not be found by the routine. 
These sources were kept in the catalog, but not considered for the 
determination of the LF. 

For the study of the faint end of the LF, the number counts
of dSph candidates have to be completeness corrected. Therefore, we randomly
distributed 2000 simulated dSphs (in 100 runs) in each of
the 14 CCD fields. The magnitudes and central surface brightnesses were chosen
such that they extended well beyond the observed parameter space 
at the faint limits. The 
optimized detection parameters were used to recover the artificial galaxies.
The same selection criteria as for the discovered dSphs were applied to
derive the completeness values as a function of magnitude and central surface
brightness (see Fig.~\ref{mumag}). For the number counts of the LF
the completeness in each CCD field has been corrected individually to
account for the differing number densities.

Further details of the observations and data analysis will be given in Hilker 
et al. (2003, in prep.).

\section{Dwarf spheroidal candidates in Fornax}

Since the faintest Local Group dSphs appear just resolved when projected 
to the Fornax distance ($1\arcsec$ corresponds to about 92 pc at 19 Mpc;
Ferrarese et al. \cite{ferra00}), the first selection criterion for dSphs 
in Fornax was a measured FWHM larger than $5\arcsec$.
Furthermore the color range was restricted to within $\pm2\sigma$ of the 
color-magnitude relation (see Fig.~\ref{cmd}). Finally, probable dSphs 
candidates should be located within $\pm2\sigma$ of the magnitude-surface 
brightness relation (Fig.~\ref{mumag}).

The photometric parameters of the dwarf galaxies have been derived from the
analysis of their surface brightness profiles: the total magnitude by a
curve of growth analysis, the color within an aperture of $8\arcsec$ diameter, 
and the central surface brightness from an exponential fit to the outer part
of the profile.

\subsection{The color-magnitude relation}

\begin{figure}
\psfig{figure=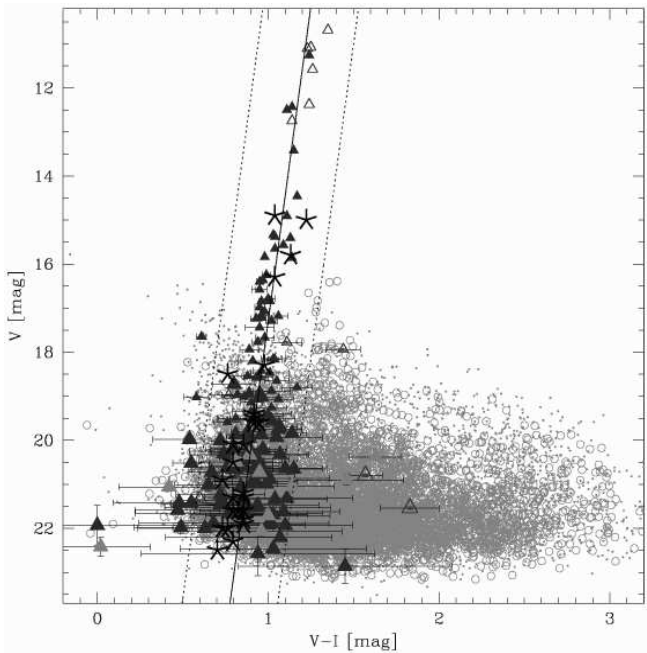,height=8.6cm,width=8.6cm
,bbllx=9mm,bblly=65mm,bburx=195mm,bbury=246mm}
\vspace{0.4cm}
\caption{\label{cmd} CMD of all extended (circles) and point sources (dots) in 
our Fornax field also in comparison with Local group dSphs. 
The formerly known Fornax members are marked as
small triangles. The larger triangles are newly discovered dSphs. Light
grey triangles are dwarfs that have been detected only by eye.
Open symbols mark
galaxies that lie outside $2\sigma$ of the maginitude-surface brightness
relation.
The solid line is a fit to the color-magnitude relation of dEs
between $0.3<(V-I)<1.4$. Dotted lines are the $2\sigma$ deviations
from the fit. Asterisks are the Local Group dSphs (data from Grebel et al. 
\cite{greb02a}) projected to the Fornax distance.
}
\end{figure}

\begin{figure}
\psfig{figure=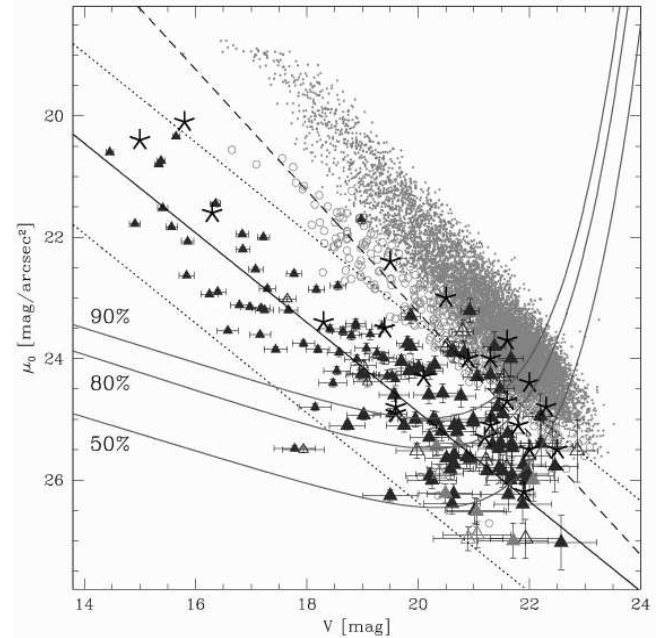,height=8.6cm,width=8.6cm
,bbllx=9mm,bblly=65mm,bburx=195mm,bbury=246mm}
\vspace{0.4cm}
\caption{\label{mumag} Magnitude-surface brightness diagram in our Fornax 
field and the Local Group dEs and dSphs. The central surface brightness 
derived from an exponential fit is plotted. Symbols are as 
in Fig.~\ref{cmd}, except that open triangles here mark
galaxies that lie outside $2\sigma$ of the color-magnitude relation.
The dashed line indicates a scale length of $2\farcs5$ for an exponential
profile. The solid and dotted lines are the fit to the magnitude-surface
brightness relation and its $2\sigma$ deviations. The solid curves
show the completeness limits of 90, 80 and 50\%.
}
\end{figure}

In Fig.~\ref{cmd} the color magnitude diagram (CMD) of all objects is shown. 
The newly discovered dSphs are highlighted by large triangles. Except for
some outliers, they follow a well defined 
color-magnitude sequence in the sense that the fainter galaxies are bluer.
A linear fit to the data yields:
$(V-I) = -0.035\cdot V + 1.61$ with a rms of 0.14. 

The color-magnitude relation of early-type galaxies is well known from other
clusters (e.g. Coma cluster, Secker et al. \cite{seck97}). It is likely
explained by a strong metallicity-luminosity relation (see Poggianti et al.
\cite{pogg01a}). Here we show for the first time that this
relation extends all the way down to the regime of dSphs.
For the Local Group dSphs, there do not exist homogeneous $(V-I)$
colors. However, assuming that they are single stellar populations and then
transforming their average iron abundances [Fe/H] (Grebel et al. \cite{greb02a})
to $(V-I)$ colors using equation (4) given in Kissler-Patig et al.
(\cite{kiss98a}), they follow surprisingly well the same
color-magnitude relation (Fig.~\ref{cmd}).

\subsection{The magnitude-surface brightness relation}

Dwarf ellipticals are known to follow a $r_{\rm eff}$-$M_V$ relation distinct
to that of giant ellipticals (e.g. Bender et al. \cite{bend92}).
Also, they follow a tight $M_V$--$\mu_V$ relation in the sense that central
surface brightness increases with increasing luminosity (Ferguson \&
Sandage \cite{ferg88}). The validity of this relation has been
a subject of lively debate. A number of authors have
argued against the existence of a magnitude-surface brightness relation for 
dEs (i.e. Irwin et al. \cite{irwi90}) and questioned the cluster membership 
assignement to dEs based on morphology. However, Drinkwater et al. 
(\cite{drin01b}) confirm the surface brightness-magnitude relation for Fornax 
dwarfs, based on their spectroscopic survey.

Our data shows that the magnitude-surface brightness relation continues
even to fainter magnitudes. As one can see in Fig.~\ref{mumag}, the sequence
of Fornax cluster dSphs matches well the location of Local Group
dSphs in this plot (data from Grebel et al. \cite{greb02a}). Note that some
more compact dSphs still might be hidden in the barely resolved objects.

\section{The faint end of the luminosity function}

\begin{figure}
\psfig{figure=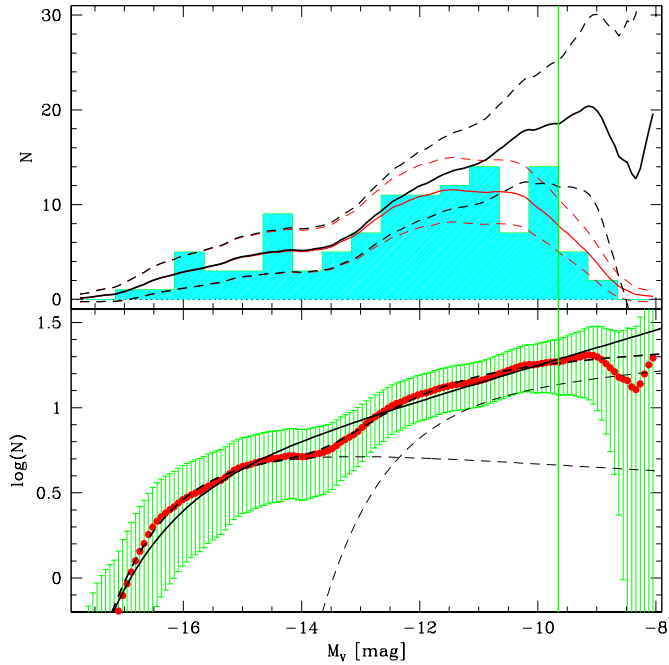,height=8.6cm,width=8.6cm
,bbllx=9mm,bblly=65mm,bburx=195mm,bbury=246mm}
\vspace{0.4cm}
\caption{\label{lkf} Luminosity function of dEs and dSphs in
Fornax. The 50\% completeness limit is indicated by a vertical line.
Upper panel: The shaded histogram are the uncorrected number counts. The thin
line gives a binning independent representation of the counts (Epanechnikov
kernel of 0.5 mag width). The thick line are the completeness corrected counts.
with the $1\sigma$ uncertainty limits (dashed).
Lower panel: completeness corrected number counts in logarithmic
representation. The best fitting single Schechter function (solid curve) and
sum of two Schechter functions (dashed curves) are shown.
}
\end{figure}

In Fig.~\ref{lkf}, the luminosity distribution of the dEs and dSphs in Fornax
is shown.  When fitting a Schechter (\cite{sche76}) 
function to the counts with a completeness larger than 50\% the
faint-end slope is $\alpha = -1.11\pm0.10$. This agrees with the result by
Ferguson \& Sandage (\cite{ferg88}), although their faintest
dwarfs were 2.5 mag brighter than ours.

There seems to be a dip in the luminosity distribution at about $M_V=-14$ mag.
Although this might be due to small number counts (the amplitude of the dip
is about equal to the Poisson error of the number density at the corresponding
magnitude), it is interesting to note
that this is near the luminosity where the separation of dEs 
and dSphs is defined (e.g. Grebel \cite{greb01}). 
A sum of two Schechter functions does better fit
the data (see lower panel in Fig.~\ref{lkf}) but does not
change the faint -nd slope.

\subsection{Comparison with Kambas' results}

Recently, Kambas et al. (\cite{kamb00}) reported the discovery of a very large
number ($>3400$) of very low surface brightness (VLSB) galaxies which are 
concentrated towards the center of the Fornax cluster. The resulting faint-end 
slope is very steep: $\alpha\simeq-2$. Thus, although having a comparable 
limiting magnitude, Kambas' results are in large disagreement with ours. 
The data sets and selection criteria of both studies are quite different.
Kambas et al. took $R$ band images on a CCD chip with a scale of $2\farcs3$ 
per pixel, thus having a 3 times lower resolution than we do. The VLSB 
candidates were selected by lower limits in the scale length and surface 
brightness derived from the isophotal magnitude and area measured by 
SExtractor. No curve-of-growth analysis or surface brightness profile fitting
was performed to cross-check the SExtractor results.

To find the reason for this discrepancy, Kambas kindly provided
us with his data. In the overlapping region of 
both data sets, all VLSBs (in total 473) were also detected by us.
According to Kambas et al., only 110 of them are background objects.
However, by comparison with our data we found that at least 300 of the 473 
objects are clear non-members of the Fornax clusters. About 90 objects 
are clear point sources even at our 3 times higher resolution. More
than 250 objects have a $(V-I)$ color redder than 1.4 mag, and thus are 
background galaxies (which could not be detected with only one filter at hand). 
Many other VLSB candidates are either clearly multiple or are
located in the halo of brighter galaxies or close to saturated stars.

We therefore suggest that most of Kambas' VLSB dwarf candidates are not dwarf 
galaxies in Fornax. And we conclude that SExtractor values that characterize 
the sizes of objects do not provide reliable selection criteria close to the 
resolution limit and close to the magnitude and surface brightness limit.


\section{Summary and Conclusions}

A deep wide field survey of the central 2.4 square degrees of the Fornax 
cluster has revealed a large population of previously undetected low surface 
brightness dwarf galaxies whose brightness profiles could been measured. 
It is found that they resemble the Local Group dwarf
spheroidals. In particular, they follow the 
same magnitude-surface brightness relation. Also they follow a quite tight
color-magnitude relation which can be explained by a metallicity effect, in the
sense that fainter dwarf galaxies are more metal-poor than brighter ones.

Our study has therefore shown for the first time that the counterparts of the 
Local Group dwarf spheroidals 
do exist in cluster environments. The faint end slope of their luminosity
function is flat ($\alpha=-1.1\pm0.1$) in comparison to the expected initial 
slope of small 
halos in current CDM simulations of galaxy clusters. If theory is right, this 
might point to the destruction of a large number of dwarfs during the evolution
of a cluster. One might suggest that the debris of these dwarfs have partly 
built up the huge cD halo around the central galaxy (e.g. Hilker et al. 
\cite{hilk99c}).

A dip in the luminosity function at the transition between dwarf ellipticals
and dwarf spheroidals (at $M_V\simeq -14$ mag) might point to two distinct 
families of galaxies.
Although their photometric properties are very similar, dEs and dSphs might
have experienced different evolutionary histories, i.e. that dSphs evolved from
dwarf irregulars that have consumed their gas or have lost it by ram pressure
stripping (Grebel \cite{greb01}).

Our study has also shown that the combination of deep multi-color photometry 
in a wide field with a sufficient resolution is crucial in order to
unambigously identify dwarf spheroidal candidates in nearby clusters. 
In upcoming studies one should push the limits to even fainter magnitudes
and surface brightnesses, since we have still not seen a mass cutoff of the
smallest galactic systems.


\acknowledgements
We gratefully thank A. Kambas for providing us with his data. LI and SM 
acknowlegde support from `Proyecto FONDAP 15010003'. SM was supported by
DAAD PhD grant Kennziffer D/01/35298. MH thanks du~Pont staff for support.

\enddocument